\documentclass[11pt, a4paper]{article}
\usepackage[utf8]{inputenc}
\usepackage[T1]{fontenc}
\pdfoutput = 1
\usepackage{jheppub}

\usepackage{graphicx}
\usepackage{listings}
\usepackage{amsthm}
\usepackage{amsmath}
\usepackage{amssymb}
\usepackage{amsfonts}
\usepackage{braket}
\usepackage{bbm}
\usepackage{bbold}
\usepackage{float}
\usepackage{mathabx}
\usepackage{url}
\usepackage{upgreek}
\usepackage[normalem]{ulem}

\usepackage{mathtools}
\allowdisplaybreaks
\numberwithin{equation}{section}

\usepackage{color}
    \definecolor{darkgreen}{rgb}{0,0.5,0}
    \definecolor{darkred}{rgb}{0.5,0,0}
    \definecolor{darkblue}{rgb}{0,0,0.6}
    \definecolor{purple}{rgb}{0.4,.2,0.7}

\usepackage[ragged]{footmisc}
\usepackage{dsfont}

\newcommand{\dd}{{\mathrm{d}}}
\newcommand{\Tr}{{\mathrm{Tr}}}
\newcommand{\Str}{{\mathrm{Str}}}
\newcommand{\Ber}{{\mathrm{Ber}}}
\newcommand{\VdM}{{\mathrm{VdM}}}
\newcommand{\sgn}{{\mathrm{sgn}}}
\newcommand{\Res}{{\mathrm{Res}}}
\usepackage{mathrsfs}

\author{Dan Stefan Eniceicu}
\affiliation{Stanford Institute for Theoretical Physics, Stanford University,\\Stanford, CA 94305, USA}
\emailAdd{eniceicu@stanford.edu}
\title{\boldmath Comments on the Giant-Graviton Expansion of the Superconformal Index}
\abstract{Work by Gaiotto and Lee, and by Imamura and collaborators suggests that the superconformal index of $U(N)$ gauge theory should be expressible as a convergent series whose terms are indices of associated $U(k)$ gauge theories realized as the worldvolume theories of stacks of $k$ giant-graviton branes. A different expansion for the index provided by Murthy was shown to hold very generally, but the connection to the first expansion was not immediately clear. We study the relation between the two expansions and propose a prescription for extracting the terms of the first series from those of the second. We follow this prescription in the case of the $1/2$-BPS index and show that the contribution of the $m$th term of the first expansion is fully encoded in the first $m$ terms of the second. In addition, we identify the $m$th term of the second expansion with the expectation value of the $N$th power of the superdeterminant in a $U(m|m)$ superunitary matrix integral, which hints at a brane/anti-brane origin for the term.}

\begin{document}
\maketitle
\flushbottom
\section{Introduction and Outline of the Main Results} \label{sec:intro}
To describe similarities and differences between mathematical objects of the same type, one often defines quantities which remain unchanged under particular types of deformations of the objects. These quantities are called invariants and are crucial to the characterization of relations between objects in the same category. In a supersymmetric quantum theory with Hilbert space $\mathcal{H}$, a celebrated invariant is the Witten index \cite{Witten} which is defined by replacing the trace in the standard partition function with the supertrace,
\begin{equation}
    I_{W}(\beta)=\Tr_{\mathcal{H}}\left[(-1)^Fe^{-\beta H}\right],
\end{equation}
and is equal to the difference between the number of bosonic and fermionic ground states. The Witten index has the property that it remains unchanged as the strengths of the interactions of the theory are continuously varied, as long as supersymmetry is preserved during this deformation.\\\\
An important class of invariants in field theory are superconformal indices \cite{SCIndex1, SCIndex2, SCIndex3}. Loosely speaking, these are refinements of the Witten index for superconformal field theories in radial quantization. They receive contributions solely from the subset of BPS representations which do not combine into generic representations of the superconformal algebra under continuous changes of the parameters of the theory which preserve the superconformal symmetry, and they encode all information about protected states obtainable solely from group theory. Due to the invariance of the superconformal index under the continuous change in the parameters of the theory, the index becomes an important tool for probing aspects of dualities such as the AdS/CFT correspondence.\footnote{For an introduction to the superconformal index, see for instance \cite{SCIndexIntro}.}\\\\
Recent work by Imamura and collaborators \cite{Imamura1, Imamura2, Imamura3, Imamura4, Imamura5, Imamura6, Imamura7}, and by Gaiotto and Lee \cite{GL} has led to the remarkable conjecture that for a variety of different gauge theories, the expression for the finite-$N$ index $Z_N(q)$ can be computed from a series of systematic corrections to the associated $N\rightarrow\infty$ expression. Concretely, in the case of four-dimensional $\mathcal{N}=4$ supersymmetric Yang-Mills theory with gauge group $U(N)$, the proposed identity for an index which depends on a single parameter $q<1$ takes the form
\begin{equation}
    \frac{Z_N(q)}{Z_{\infty}(q)}=1+\sum_{k=1}^{\infty}q^{kN}\hat{Z}_k(q). \label{eqn:GL}
\end{equation}
The left side is the ratio of the index of the $U(N)$ gauge theory and the limit of that index as $N\rightarrow\infty$, while the right side involves terms $\hat{Z}_k(q)$ which correspond to indices of associated $U(k)$ gauge theories, and have no $N$-dependence. Since the $U(N)$ gauge theory admits a dual holographic description, a natural question is whether the expression (\ref{eqn:GL}) itself admits a holographic interpretation. The aforementioned papers identify the term $\hat{Z}_k(q)$ as the index of the $U(k)$ worldvolume gauge theory of a stack of $k$ ``giant graviton'' D3-branes in the dual string theory, hence the name ``giant-graviton expansion'' for (\ref{eqn:GL}). Since the publication of these results, superconformal indices and giant gravitons have been the focus of an impressive collection of subsequent works \cite{Lee, SCIUpdate1, SCIUpdate20, SCIUpdate2, SCIUpdate3, Murthy, SCIUpdate4, SCIUpdate5, SCIUpdate6, SCIUpdate7, SCIUpdate8, SCIUpdate9, SCIUpdate10, SCIUpdate11, SCIUpdate12, SCIUpdate13, SCIUpdate14, SCIUpdate15, SCIUpdate16, SCIUpdate17, SCIUpdate18, LiuRajappa, SCIUpdate19}.\\\\
Following a very different route \cite{Murthy}, Murthy showed that by writing the $U(N)$ gauge theory index as a unitary matrix integral in terms of the single-letter index $f(q)$,
\begin{equation}
    Z_N(q)=\int_{U(N)}\dd U\;\exp\left[\sum_{k=1}^\infty\frac{f\left(q^k\right)}{k}\Tr\left(U^k\right)\Tr\left(U^{-k}\right)\right],
\end{equation}
and applying a theorem of Geronimo and Case \cite{GC}, later rediscovered by Borodin and Okounkov \cite{BO}, one can write a convergent expansion for the index. This expansion takes the form
\begin{equation}
    \frac{Z_N(q)}{Z_{\infty}(q)}=1+\sum_{m=1}^{\infty}G_N^{(m)}(q). \label{eqn:Murthy}
\end{equation}
Murthy showed that if the lowest power of $q$ appearing in the single-letter index $f(q)$ was $q^{\alpha}$, then each of the terms $G_N^{(m)}(q)$ in the expansion would be a power series in $q$ with lowest power of $q$ larger or equal to $q^{\alpha m N}$. In particular, for $\alpha=1$, this means that the $m$th term in Murthy's expansion only starts contributing at order $q^{mN}$ or higher, which is also the case for $q^{mN}\hat{Z}_m(q)$ in (\ref{eqn:GL}). While Murthy's expansion is a rigorously proven identity, the individual terms $G_N^{(m)}(q)$ don't yet have a sharp interpretation like the terms $\hat{Z}_m(q)$ in Gaiotto and Lee's expansion do, to the extent of our knowledge. In fact, it is also not immediately clear what the connection between the two expansions is, aside from the fact that the two expansions clearly differ \cite{LiuRajappa}.\\\\
The goal of our work is two-fold. The first goal is to propose such a connection, which we study in the particularly simple case of the index which counts $1/2$-BPS operators. We believe the analysis should generalize to other indices. The second goal is to provide a mathematical interpretation of the terms $G_N^{(m)}(q)$ as expectation values of the $N$th power of the superdeterminant (or Berezinian) in an associated $U(m|m)$ superunitary matrix integral. Such integrals (more precisely, their Hermitian $U(n|m)$-symmetric analogs) have appeared previously in the literature in connection with brane/anti-brane systems.\\\\
The main idea behind our proposal is to study the generating function of $U(N)$ indices,
\begin{equation}
    \mathcal{Z}(\zeta;q)=\sum_{N=1}^{\infty}\zeta^NZ_N(q),\label{eqn:GrandGen}
\end{equation}
in both expansions. For the giant-graviton expansion\footnote{Throughout this work, we will use the term ``giant-graviton expansion'' to refer exclusively to the expansion (\ref{eqn:GL}), where $\hat{Z}_k(q)$ is $N$-independent.},
\begin{equation}
    \frac{\mathcal{Z}(\zeta;q)}{Z_\infty(q)}=\sum_{N=1}^{\infty}\zeta^N\left(1+\sum_{k=1}^{\infty}q^{Nk}\hat{Z}_k(q)\right)=\frac{\zeta}{1-\zeta}+\sum_{k=1}^\infty\frac{\zeta q^k}{1-\zeta q^k}\hat{Z}_k(q),\label{eqn:GrandGenGL}
\end{equation}
which means that as a function of $\zeta$, the ratio $\mathcal{Z}(\zeta;q)/Z_\infty(q)$ is a meromorphic function of $\zeta$, with simple poles at $\zeta_k=q^{-k}$, and residues $-\hat{Z}_k(q)$ for $k\in\mathbb{N}^*$. On the other hand, as we explain later in the paper, one can obtain an expression for the terms in (\ref{eqn:Murthy}),
\begin{align}
    G_N^{(m)}(q)&=\frac{(-1)^m}{\left(m!\right)^2}\prod_{j=1}^m\oint\frac{\dd u_j}{2\pi i}\prod_{i=1}^m\oint\frac{\dd v_i}{2\pi i v_i^2}\Bigg\{\left[\prod_{i=1}^m\left(\frac{u_i}{v_i}\right)\right]^N\left[\det\left(\frac{1}{1-u_i/v_j}\right)_{i,j=1,\dots,m}\right]^2\cdot\nonumber\\
    &\cdot\prod_{i,j=1}^m\prod_{l=1}^\infty\left[\frac{\left(1-q^lu_i/u_j\right)\left(1-q^lv_i/v_j\right)}{\left(1-q^lu_i/v_j\right)\left(1-q^lv_i/u_j\right)}\right]^{\hat{a}_l}\Bigg\}.\label{eqn:MurthysGiant}
\end{align}
We will specify the choice of contours of integration as well as the relation of the exponents $\hat{a}_l$ to the single-letter index $f(q)$ in the upcoming sections. Combining the previous expression with (\ref{eqn:Murthy}) and (\ref{eqn:GrandGen}), and assuming the infinite sums commute, we find:
\begin{align}
    \frac{\mathcal{Z}(\zeta;q)}{Z_\infty(q)}&=\frac{\zeta}{1-\zeta}+\sum_{m=1}^\infty\frac{(-1)^m}{\left(m!\right)^2}\prod_{j=1}^m\oint\frac{\dd u_j}{2\pi i}\prod_{i=1}^m\oint\frac{\dd v_i}{2\pi i v_i^2}\Bigg\{\frac{\zeta\prod_{i=1}^m\left(u_i/v_i\right)}{1-\zeta\prod_{i=1}^m\left(u_i/v_i\right)}\cdot\nonumber\\&\cdot\left[\det\left(\frac{1}{1-u_i/v_j}\right)_{i,j=1,\dots,m}\right]^2\prod_{i,j=1}^m\prod_{l=1}^\infty\left[\frac{\left(1-q^lu_i/u_j\right)\left(1-q^lv_i/v_j\right)}{\left(1-q^lu_i/v_j\right)\left(1-q^lv_i/u_j\right)}\right]^{\hat{a}_l}\Bigg\}.\label{eqn:GrandGenGrav}
\end{align}
The point we want to emphasize is that the contour integrals will pick out a particular sum of residues as $u_i,v_i\rightarrow0$, with the product $\prod_{i=1}^m\left(u_i/v_i\right)\rightarrow q^k$ for various $k\in\mathbb{N}^*$,
\begin{equation}
    \frac{\mathcal{Z}(\zeta;q)}{Z_\infty(q)}=\frac{\zeta}{1-\zeta}+\sum_{k=1}^\infty\frac{\zeta q^k}{1-\zeta q^k}c_k(q),
\end{equation}
where the coefficients $c_k(q)$ have no $\zeta$-dependence. These coefficients can then be identified with those in (\ref{eqn:GrandGenGL}), $\hat{Z}_k(q)=c_k(q)$, hence allowing us to extract the $k$th giant-graviton contribution from Murthy's expansion.\footnote{See, however, the discussion in section~\ref{sec:conclusions} regarding the potential appearance of higher-order poles in $\zeta$ as we perform the integrals. This possibility will not be relevant for the case of the $1/2$-BPS index which we analyze in this work.}\\\\
To summarize, if we expect a relation such as (\ref{eqn:GL}) to hold, we can obtain the individual giant-graviton contributions $\hat{Z}_k(q)$ from Murthy's expansion (\ref{eqn:Murthy}) by combining the finite-$N$ indices into the generating function (\ref{eqn:GrandGen}) and performing the contour integrals in (\ref{eqn:GrandGenGrav}). In fact, we can also refrain from performing the sum over $m$ in (\ref{eqn:GrandGenGrav}) and see explicitly the contribution of Murthy's $m$th term $G_N^{(m)}(q)$ to each of the terms $\hat{Z}_k(q)$ in the giant-graviton expansion.\\\\
In the particularly simple specialization of the index, which counts $1/2$-BPS operators, the single-letter index is given by $f(q)=q$, and the exponents $\hat{a}_l$ appearing in (\ref{eqn:MurthysGiant}) and (\ref{eqn:GrandGenGrav}) are all equal to 1. In this case, the finite-$N$ index takes the form,
\begin{equation}
    Z_N^{(1/2)}(q)=\prod_{n=1}^N\frac{1}{1-q^n},
\end{equation}
while Gaiotto and Lee's $k$th giant-graviton contribution is
\begin{equation}
    \hat{Z}_k^{(1/2)}(q)=(-1)^kq^{k(k+1)/2}\prod_{n=1}^k\frac{1}{1-q^n}=\prod_{n=1}^k\frac{1}{1-q^{-n}}. \label{eqn:knownZkhat}
\end{equation}
As we shall see in the coming sections, each of Murthy's terms $G_N^{(m)}(q)$ will contain a part of $\hat{Z}_k(q)$ for $k\geq m$. For instance, Murthy's first two terms can be expressed as
\begin{align}
    G_N^{(1)}(q)&=\sum_{k=1}^\infty(-1)^kq^{Nk+k(k+1)/2}\frac{1}{1-q^k}\nonumber\\
    &=q^N\left(\frac{-q}{1-q}\right)+q^{2N}\left(\frac{q^3}{1-q^2}\right)+q^{3N}\left(\frac{-q^6}{1-q^3}\right)+\dots,\\
    G_N^{(2)}(q)&=\frac{1}{2}\sum_{k=2}^\infty(-1)^kq^{Nk+k(k+1)/2}\left[-\frac{k-1}{1-q^k}+\sum_{l=1}^{k-1}\frac{1}{\left(1-q^l\right)\left(1-q^{k-l}\right)}\right]\nonumber\\
    &=q^{2N}\left(\frac{q^4}{(1-q)\left(1-q^2\right)}\right)+q^{3N}\left(\frac{-q^7\left(1+2q\right)}{\left(1-q\right)\left(1-q^2\right)\left(1+q+q^2\right)}\right)+\dots,
\end{align}
which allows us to extract the first two giant-graviton contributions:
\begin{align}
    \hat{Z}_1(q)&=\frac{-q}{1-q}=\frac{1}{1-q^{-1}},\\
    \hat{Z}_2(q)&=\frac{q^3}{1-q^2}+\frac{q^4}{(1-q)\left(1-q^2\right)}=\frac{1}{\left(1-q^{-1}\right)\left(1-q^{-2}\right)}.
\end{align}
To extract $\hat{Z}_k(q)$, we will need the expressions for the first $k$ terms in Murthy's expansion, $G_N^{(1)}(q),G_N^{(2)}(q),\dots,G_N^{(k)}(q)$. As we shall see in the following sections, in the $1/2$-BPS case, these can be written as
\begin{equation}
    G_N^{(m)}(q)=\frac{1}{m!}\sum_{k=m}^\infty(-1)^kq^{Nk+k(k+1)/2}\sum_{\substack{L_1,\dots,L_m\geq1\\L_1+\dots+L_m=k}}F_{L_1,\dots,L_m}^{(m)}(q),\label{eqn:FinalExpHalfBPS}
\end{equation}
where the terms $F_{L_1,\dots,L_m}^{(m)}(q)$ satisfy the following recurrence relation for $m\geq2$,
\begin{equation}
    F_{L_1,\dots,L_m}^{(m)}(q)=\frac{1}{1-q^{L_m}}F_{L_1,\dots,L_{m-1}}^{(m-1)}(q)-\sum_{j=1}^{m-1}F_{L_1,\dots,L_{j-1},L_j+L_m,L_{j+1},\dots,L_{m-1}}^{(m-1)}(q),\label{eqn:RecurrenceHalfBPS}
\end{equation}
with
\begin{equation}
    F_{L_1}^{(1)}(q)=\frac{1}{1-q^{L_1}}.\label{eqn:FirstTermHalfBPS}
\end{equation}
One can immediately read off the prediction for the giant-graviton contribution from (\ref{eqn:FinalExpHalfBPS}),
\begin{equation}
    \hat{Z}_k^{(1/2)}(q)=(-1)^kq^{k(k+1)/2}\sum_{m=1}^k\frac{1}{m!}\sum_{\substack{L_1,\dots,L_m\geq1\\L_1+\dots+L_m=k}}F_{L_1,\dots,L_m}^{(m)}(q). \label{eqn:foundZkhat}
\end{equation}
This expression can further be proven \cite{Daniel} to reduce to the known answer for the associated $U(k)$ gauge theory index, equation (\ref{eqn:knownZkhat}), by a combinatorial argument.\\\\ 
Finally, we will also show how one can identify the $m$th term of Murthy's Fredholm determinant expansion for the $U(N)$ index as the expectation value of the $N$th power of the Berezinian of a $U(m|m)$ superunitary matrix $W$, weighted by the exponential of a sum of products of supertraces,
\begin{equation}
    G_N^{(m)}(q)=\int_{U(m|m)}\dd W\;\Ber(W)^N\exp\left[-\sum_{k=1}^\infty\frac{1}{k}\frac{f\left(q^k\right)}{1-f\left(q^k\right)}\Str\left(W^k\right)\Str\left(W^{-k}\right)\right].
\end{equation}
One might tentatively conjecture this expression points to a $U(m|m)$ supergroup gauge theory origin of the terms in the Fredholm determinant expansion, similar to how the terms in (\ref{eqn:GL}) have an interpretation as indices of associated $U(k)$ gauge theories realized as worldvolume theories of stacks of $k$ giant-graviton branes.\\\\
For clarity, we choose to separate the purely mathematical derivation of our two results from our discussions on their potential physical implications. Readers interested primarily in the latter will find these in section~\ref{sec:conclusions}.
Having summarized the main ideas of our argument, we now introduce the outline of the paper:
\begin{itemize}
    \item In appendix~\ref{app:A}, we review Murthy's Fredholm determinant expansion of the superconformal index, along with the Tracy-Widom evaluation of the Fredholm determinant adapted to the present case.
    \item In section~\ref{sec:2}, we explain how to arrive at equation (\ref{eqn:GrandGenGrav}) starting from the Fredholm determinant expansion.
    \item In section~\ref{sec:3}, we perform the integrals in equation (\ref{eqn:GrandGenGrav}) for the $1/2$-BPS index to derive (\ref{eqn:FinalExpHalfBPS}--\ref{eqn:FirstTermHalfBPS}).
    \item In section~\ref{sec:results}, we list expressions for the first few terms $G_N^{(m)}(q)$ in the Fredholm determinant expansion of the $1/2$-BPS index, and check that the predictions for the giant-graviton contributions $\hat{Z}_k(q)$ extracted from the Fredholm determinant expansion match the known results.
    \item In section~\ref{sec:supermatrix}, we derive the expression for $G_N^{(m)}(q)$ as the expectation value of the $N$th power of the Berezinian of a $U(m|m)$ superunitary matrix.
    \item We conclude in section~\ref{sec:conclusions} with a summary and a discussion on the physical implications of our work.
    \item In appendix~\ref{app:B}, we include a combinatorial proof of the equality between the expressions in equations (\ref{eqn:knownZkhat}) and (\ref{eqn:foundZkhat}). We are grateful to Dongryul Kim for providing this proof and allowing us to include it in our work \cite{Daniel}.
\end{itemize}
\section{Integral Representation of $G_N^{(m)}(q)$}\label{sec:2}
Our starting point is the Fredholm determinant expansion of the superconformal index (\ref{eqn:finalAppendixExpression}), which we reproduce here for convenience, with $g_k=f\left(q^k\right)$:
\begin{align}
    Z_N\left(q\right)&=Z_\infty\left(q\right)\sum_{m=0}^\infty\frac{(-1)^m}{\left(m!\right)^2}[1]\Bigg\{\prod_{i=1}^m\left(\frac{u_i}{v_i}\right)^{N+1}\cdot\left[\det\left(\frac{1}{1-u_i/v_j}\right)_{i,j=1,\dots,m}\right]^2\cdot\nonumber\\
    &\cdot\exp\left[\sum_{k=1}^\infty\frac{1}{k}\frac{f\left(q^k\right)}{\left(1-f\left(q^k\right)\right)}\sum_{i=1}^m\left(u_i^k-v_i^k\right)\sum_{j=1}^m\left(v_j^{-k}-u_j^{-k}\right)\right]\Bigg\}, \label{eqn:startingpointforsupermatrix}
\end{align}
where
\begin{equation}
    Z_\infty(q)\equiv\prod_{l=1}^\infty\frac{1}{1-f\left(q^l\right)}
\end{equation}
represents the $N\rightarrow\infty$ limit of the index, and the notation $[1]\{\dots\}$ denotes picking out the constant coefficient from the power series in variables $u_i$, $v_j$ inside the brackets. The fraction $1/\left(1-u_i/v_j\right)$ is interpreted as the power series obtained by expanding for $\left|u_i/v_j\right|<1$.\\\\
Assuming the fraction $f/(1-f)$ can be written as a power series in $q$ with integer coefficients,
\begin{equation}
    \frac{f(q)}{1-f(q)}=\sum_{l=1}^\infty\hat{a}_lq^l,\;\;\;\hat{a}_l\in\mathbb{Z},\label{eqn:defahatl}
\end{equation}
one finds
\begin{align}
    Z_N\left(q\right)&=Z_\infty\left(q\right)\sum_{m=0}^\infty\frac{(-1)^m}{\left(m!\right)^2}[1]\Bigg\{\prod_{i=1}^m\left(\frac{u_i}{v_i}\right)^{N+1}\cdot\left[\det\left(\frac{1}{1-u_i/v_j}\right)_{i,j=1,\dots,m}\right]^2\cdot\nonumber\\
    &\cdot\prod_{i,j=1}^m\prod_{l=1}^\infty\exp\left[\hat{a}_l\sum_{k=1}^\infty\frac{1}{k}q^{lk}\left(u_i^k-v_i^k\right)\left(v_j^{-k}-u_j^{-k}\right)\right]\Bigg\}.
\end{align}
One can further identify
\begin{align}
    \exp\left[\sum_{k=1}^\infty\frac{1}{k}q^{lk}u_i^kv_j^{-k}\right]&=\exp\left[-\log\left(1-q^lu_i/v_j\right)\right]\nonumber\\
    &=\frac{1}{\left(1-q^lu_i/v_j\right)},
\end{align}
provided both sides are expanded assuming $\left|u_i/v_j\right|<|q|^{-1}$. Similarly, one can identify
\begin{align}
    \exp\left[\sum_{k=1}^\infty\frac{1}{k}q^{lk}\left(u_i^k-v_i^k\right)\left(v_j^{-k}-u_j^{-k}\right)\right]=\frac{\left(1-q^lu_i/u_j\right)\left(1-q^lv_i/v_j\right)}{\left(1-q^lu_i/v_j\right)\left(1-q^lv_i/u_j\right)},
\end{align}
provided both sides are expanded assuming $\left|u_i/v_j\right|<|q|^{-1}$, $\left|v_i/u_j\right|<|q|^{-1}$, $\left|u_i/u_j\right|<|q|^{-1}$, and $\left|v_i/v_j\right|<|q|^{-1}$.\\\\
This allows us to write
\begin{align}
    Z_N\left(q\right)&=Z_\infty\left(q\right)\sum_{m=0}^\infty\frac{(-1)^m}{\left(m!\right)^2}[1]\Bigg\{\prod_{i=1}^m\left(\frac{u_i}{v_i}\right)^{N+1}\cdot\left[\det\left(\frac{1}{1-u_i/v_j}\right)_{i,j=1,\dots,m}\right]^2\cdot\nonumber\\
    &\cdot\prod_{i,j=1}^m\prod_{l=1}^\infty\left[\frac{\left(1-q^lu_i/u_j\right)\left(1-q^lv_i/v_j\right)}{\left(1-q^lu_i/v_j\right)\left(1-q^lv_i/u_j\right)}\right]^{\hat{a}_l}\Bigg\},
\end{align}
provided $|q|<\left|u_i/v_j\right|<1$, $|q|<\left|u_i/u_j\right|<|q|^{-1}$, and $|q|<\left|v_i/v_j\right|<|q|^{-1}$ for all $1\leq i,j\leq m$.\\\\
So far, the expressions we have been working with have been formal algebraic expressions over appropriate rings of power series in $u_j$ and $v_i$. However, the previous expression has an analytic analog which is obtained by replacing $[1]\{\dots\}$ with the corresponding contour integrals which pick out the constant coefficient of the Laurent series in $u_j$, $v_i$, now defined as complex variables. Concretely,
\begin{align}
    Z_N(q)&=Z_\infty(q)\sum_{m=0}^\infty\frac{(-1)^m}{\left(m!\right)^2}\prod_{j=1}^m\oint\frac{\dd u_j}{2\pi i u_j}\prod_{i=1}^m\oint\frac{\dd v_i}{2\pi i v_i}\Bigg\{\prod_{i=1}^m\left(\frac{u_i}{v_i}\right)^{N+1}\cdot\nonumber\\
    &\cdot\left[\det\left(\frac{1}{1-u_i/v_j}\right)_{i,j=1,\dots,m}\right]^2\prod_{i,j=1}^m\prod_{l=1}^\infty\left[\frac{\left(1-q^lu_i/u_j\right)\left(1-q^lv_i/v_j\right)}{\left(1-q^lu_i/v_j\right)\left(1-q^lv_i/u_j\right)}\right]^{\hat{a}_l}\Bigg\},
\end{align}
where we are allowed to pick the contours such that $|q|<\left|u_i/v_j\right|<1$, $|q|<\left|u_i/u_j\right|<|q|^{-1}$, and $|q|<\left|v_i/v_j\right|<|q|^{-1}$ for all $1\leq i,j\leq m$.\\\\
Thus, the $m$th term in Murthy's Fredholm determinant expansion takes the form
\begin{align}
    G_N^{(m)}(q)&=\frac{(-1)^m}{\left(m!\right)^2}\prod_{j=1}^m\oint\frac{\dd u_j}{2\pi i}\prod_{i=1}^m\oint\frac{\dd v_i}{2\pi i v_i^2}\Bigg\{\left[\prod_{i=1}^m\left(\frac{u_i}{v_i}\right)\right]^{N}\left[\det\left(\frac{1}{1-u_i/v_j}\right)_{i,j=1,\dots,m}\right]^2\cdot\nonumber\\
    &\cdot\prod_{i,j=1}^m\prod_{l=1}^\infty\left[\frac{\left(1-q^lu_i/u_j\right)\left(1-q^lv_i/v_j\right)}{\left(1-q^lu_i/v_j\right)\left(1-q^lv_i/u_j\right)}\right]^{\hat{a}_l}\Bigg\},
\end{align}
where the contours can be chosen such that the integrated variables satisfy the bounds specified previously. For instance, one can choose the contours for all the $u_j$ to be circles of some given radius $r$, centered at the origin, and the contours for all the $v_i$ to be circles of some given radius $R$, centered at the origin, with $|q|<r/R<1$. Multiplying by $\zeta^N$ and performing the sum over $N$ and $m$, we find (\ref{eqn:GrandGenGrav}),
\begin{align}
    \frac{\mathcal{Z}(\zeta;q)}{Z_\infty(q)}&=\frac{\zeta}{1-\zeta}+\sum_{m=1}^\infty\frac{(-1)^m}{\left(m!\right)^2}\prod_{j=1}^m\oint\frac{\dd u_j}{2\pi i}\prod_{i=1}^m\oint\frac{\dd v_i}{2\pi i v_i^2}\Bigg\{\frac{\zeta\prod_{i=1}^m\left(u_i/v_i\right)}{1-\zeta\prod_{i=1}^m\left(u_i/v_i\right)}\cdot\nonumber\\&\cdot\left[\det\left(\frac{1}{1-u_i/v_j}\right)_{i,j=1,\dots,m}\right]^2\prod_{i,j=1}^m\prod_{l=1}^\infty\left[\frac{\left(1-q^lu_i/u_j\right)\left(1-q^lv_i/v_j\right)}{\left(1-q^lu_i/v_j\right)\left(1-q^lv_i/u_j\right)}\right]^{\hat{a}_l}\Bigg\}.\label{eqn:210}
\end{align}
The similarity of this equation, which expresses the generating function of $U(N)$ indices in terms of the Fredholm determinant expansion, and equation (\ref{eqn:GrandGenGL}),
\begin{equation}
    \frac{\mathcal{Z}(\zeta;q)}{Z_\infty(q)}=\frac{\zeta}{1-\zeta}+\sum_{k=1}^\infty\frac{\zeta q^k}{1-\zeta q^k}\hat{Z}_k(q),
\end{equation}
as functions of $\zeta$ is the motivation for asking whether it is possible to extract the giant-graviton contributions $\hat{Z}_k$ from (\ref{eqn:GrandGenGL}) using residue arguments. We answer this question in the affirmative in the next section for the case of the $1/2$-BPS index.
\section{Expansion of the $1/2$-BPS Index}\label{sec:3}
For the $1/2$-BPS case, the single-letter index is given by $f(q)=q$. Hence, the coefficients $\hat{a}_l$ defined in (\ref{eqn:defahatl}),
\begin{equation*}
    \frac{f(q)}{1-f(q)}=\sum_{l=1}^\infty\hat{a}_lq^l
\end{equation*}
will all be equal to $1$.\\\\
Our starting point is equation (\ref{eqn:GrandGenGrav}), the integral representation of the Fredholm determinant expansion for the generating function of $U(N)$ indices. In fact, we will refrain from performing the summation over $m$ and instead focus on individual terms in the expansion. Letting
\begin{equation}
    A_m(\zeta;q)\equiv\sum_{N=1}^\infty \zeta^N G_N^{(m)}(q),\label{eqn:Amdef}
\end{equation}
the generating function satisfies
\begin{equation}
    \frac{\zeta}{1-\zeta}+\sum_{m=1}^\infty A_m(\zeta;q)=\frac{\mathcal{Z}(\zeta;q)}{Z_\infty(q)}=\frac{\zeta}{1-\zeta}+\sum_{k=1}^\infty\frac{\zeta q^k}{1-\zeta q^k}\hat{Z}_k(q). \label{eqn:ConnectingZk}
\end{equation}
In the case of the $1/2$-BPS index, we can read off the following integral representation for the individual terms from (\ref{eqn:210}) with $\hat{a}_l=1$,
\begin{align}
    A_m(\zeta;q)&=\frac{(-1)^m}{\left(m!\right)^2}\prod_{j=1}^m\oint\frac{\dd u_j}{2\pi i}\prod_{i=1}^m\oint\frac{\dd v_i}{2\pi i v_i^2}\Bigg\{\frac{\zeta\prod_{i=1}^m\left(u_i/v_i\right)}{1-\zeta\prod_{i=1}^m\left(u_i/v_i\right)}\cdot\nonumber\\&\cdot\left[\det\left(\frac{1}{1-u_i/v_j}\right)_{i,j=1,\dots,m}\right]^2\prod_{i,j=1}^m\prod_{l=1}^\infty\left[\frac{\left(1-q^lu_i/u_j\right)\left(1-q^lv_i/v_j\right)}{\left(1-q^lu_i/v_j\right)\left(1-q^lv_i/u_j\right)}\right]\Bigg\}.\label{eqn:Am}
\end{align}
Here, all $2m$ contours are circles centered at the origin, of radius $r$ for the $u_i$ variables, and of radius $R$ for the $v_i$ variables, with $r<R<r/|q|$.\\\\
Our prescription for evaluating the contour integrals consists of two parts: First, we evaluate the integrals in the $u_i$ variables, one at a time. Each time we perform an integral in a $u_i$ variable, we pick up residues from all the poles located inside the circle of radius $r$. We eventually arrive at an expression involving contour integrals solely in the $v_i$ variables. The second part of our prescription involves proving that the residues picked up by performing an integral in a $v_i$ variable reduces the integrand to a sum of integrands of a similar form, in one less variable. This allows us to write a recurrence relation from which (\ref{eqn:FinalExpHalfBPS}--\ref{eqn:FirstTermHalfBPS}) follow.
\subsection{Performing the $u_j$ integrals}
We begin by focusing on the integral in the variable $u_m$. Since $|\zeta|<1$ and $r<R$, the $\zeta$-dependent factor and the determinant will not contribute a residue. Since $|q|<1$, the poles which will contribute will be those coming from the second denominator factor, $\left(1-q^lv_i/u_m\right)$ for all $i\in\{1,\dots,m\}$ and $l\in\mathbb{N}^*$. Since the integrand of (\ref{eqn:Am}) is symmetric in the $v_j$ variables, it is sufficient to look at the poles coming from $u_m\rightarrow q^{L_m}v_m$ for $L_m\in\mathbb{N}^*$ and multiply the resulting expression by an overall factor of $m$ to account for the symmetry. Performing the integral in $u_m$, we find
\begin{align}
    A_m(\zeta;q)&=\frac{(-1)^m}{m!(m-1)!}\prod_{j=1}^{m-1}\oint\frac{\dd u_j}{2\pi i}\prod_{i=1}^m\oint\frac{\dd v_i}{2\pi i v_i^2}\sum_{L_m=1}^\infty\Bigg\{(-1)^{L_m-1}\left(1-q^{L_m}\right)q^{L_m(L_m+1)/2}v_m\cdot\nonumber\\
    &\cdot\frac{\zeta q^{L_m}\prod_{i=1}^{m-1}\left(u_i/v_i\right)}{1-\zeta q^{L_m}\prod_{i=1}^{m-1}\left(u_i/v_i\right)}\left[\det\left(\frac{1}{1-u_i/v_j}\right)_{i,j=1,\dots,m}\right]^2\cdot\left[\prod_{i=1}^{m-1}\frac{v_m-u_i}{v_m-v_i}\right]\cdot\nonumber\\
    &\cdot\left[\prod_{i=1}^{m-1}\frac{v_i-q^{L_m}v_m}{u_i-q^{L_m}v_m}\right]\prod_{i,j=1}^{m-1}\prod_{l=1}^\infty\left[\frac{\left(1-q^lu_i/u_j\right)\left(1-q^lv_i/v_j\right)}{\left(1-q^lu_i/v_j\right)\left(1-q^lv_i/u_j\right)}\right]\Bigg\}.
\end{align}
Two crucial observations are that the integrand continues to be symmetric in the $v_j$ variables for $j\in\{1,\dots,m-1\}$, and that the apparent poles $u_i\rightarrow q^{L_m}v_m$ are canceled due to the vanishing of the determinant, since the $i$th and $m$th rows of the matrix become identical. Therefore, the previous argument once again applies for the $u_{m-1}$ integral, with residues coming solely from poles $u_{m-1}\rightarrow q^{L_{m-1}}v_{m-1}$, with the associated symmetry factor of $m-1$. The pattern we described persists, and one can evaluate all the $u_j$ integrals systematically. Letting
\begin{equation}
    k\equiv \sum_{i=1}^m L_i,\nonumber
\end{equation}
we find
\begin{align}
    &A_m(\zeta;q)=\frac{1}{m!}\sum_{k=m}^\infty(-1)^kq^{k(k+1)/2}\frac{\zeta q^k}{1-\zeta q^k}\sum_{\substack{L_1,\dots,L_m\geq1\\L_1+\dots+L_m=k}}\left[\prod_{j=1}^m\left(1-q^{L_j}\right)\right]\cdot\nonumber\\
    &\cdot\prod_{i=1}^m\oint\frac{\dd v_i}{2\pi i v_i}\Bigg\{\left[\det\left(\frac{1}{1-q^{L_i}v_i/v_j}\right)_{i,j=1,\dots,m}\right]^2\prod_{1\leq i<j\leq m}\frac{\left(v_j-q^{L_i}v_i\right)\left(v_i-q^{L_j}v_j\right)}{\left(v_j-v_i\right)\left(q^{L_i}v_i-q^{L_j}v_j\right)}\Bigg\}.\label{eqn:35}
\end{align}
One can notice that the $\zeta$-dependence has already factored out. Equations (\ref{eqn:ConnectingZk}) and (\ref{eqn:35}) allow us to identify $\hat{Z}_k(q)$ as
\begin{align}
    &\hat{Z}_k(q)=(-1)^kq^{k(k+1)/2}\sum_{m=1}^k\frac{1}{m!}\sum_{\substack{L_1,\dots,L_m\geq1\\L_1+\dots+L_m=k}}\left[\prod_{j=1}^m\left(1-q^{L_j}\right)\right]\cdot\nonumber\\
    &\cdot\prod_{i=1}^m\oint\frac{\dd v_i}{2\pi i v_i}\Bigg\{\left[\det\left(\frac{1}{1-q^{L_i}v_i/v_j}\right)_{i,j=1,\dots,m}\right]^2\prod_{1\leq i<j\leq m}\frac{\left(v_j-q^{L_i}v_i\right)\left(v_i-q^{L_j}v_j\right)}{\left(v_j-v_i\right)\left(q^{L_i}v_i-q^{L_j}v_j\right)}\Bigg\}.
\end{align}
\subsection{Performing the $v_i$ integrals}
To complete the calculation, we need to evaluate the remaining $m$ integrals in the above expression. Let
\begin{align}
    F_{L_1,\dots,L_m}^{(m)}(q)&\equiv\left[\prod_{j=1}^m\left(1-q^{L_j}\right)\right]\cdot\prod_{i=1}^m\oint\frac{\dd v_i}{2\pi i v_i}\Bigg\{\left[\det\left(\frac{1}{1-q^{L_i}v_i/v_j}\right)_{i,j=1,\dots,m}\right]^2\cdot\nonumber\\
    &\cdot\prod_{1\leq i<j\leq m}\frac{\left(v_j-q^{L_i}v_i\right)\left(v_i-q^{L_j}v_j\right)}{\left(v_j-v_i\right)\left(q^{L_i}v_i-q^{L_j}v_j\right)}\Bigg\}. \label{eqn:Fm}
\end{align}
Once again, we focus on the integral in the $m$th variable. The candidate poles are $v_m\rightarrow0$, $v_m\rightarrow q^{L_n}v_n$, $v_m\rightarrow v_i$, and $v_m\rightarrow q^{L_i-L_m}v_i$, for $1\leq i\leq m-1$. However, one realizes that $v_m\rightarrow v_i$ is not a pole since the determinant will vanish in that limit, as the $i$th and $m$th columns of the matrix will coincide. Similarly, $v_m\rightarrow q^{L_i-L_m}v_i$ is not a pole since the $i$th and $m$th rows of the matrix will coincide.\\\\
For $v_m\rightarrow0$,
\begin{equation}
    \det\left(\frac{1}{1-q^{L_i}v_i/v_j}\right)_{i,j=1,\dots,m}\rightarrow\frac{1}{1-q^{L_m}}\det\left(\frac{1}{1-q^{L_i}v_i/v_j}\right)_{i,j=1,\dots,m-1}.
\end{equation}
Therefore,
\begin{align}
    &\Res_{v_m\rightarrow0}\Bigg\{\frac{\dd v_m}{2\pi i v_m}\left[\det\left(\frac{1}{1-q^{L_i}v_i/v_j}\right)_{i,j=1,\dots,m}\right]^2\prod_{1\leq i<j\leq m}\frac{\left(v_j-q^{L_i}v_i\right)\left(v_i-q^{L_j}v_j\right)}{\left(v_j-v_i\right)\left(q^{L_i}v_i-q^{L_j}v_j\right)}\Bigg\}=\nonumber\\
    &=\frac{1}{\left(1-q^{L_m}\right)^2}\Bigg\{\left[\det\left(\frac{1}{1-q^{L_i}v_i/v_j}\right)_{i,j=1,\dots,m-1}\right]^2\prod_{1\leq i<j\leq m-1}\frac{\left(v_j-q^{L_i}v_i\right)\left(v_i-q^{L_j}v_j\right)}{\left(v_j-v_i\right)\left(q^{L_i}v_i-q^{L_j}v_j\right)}\Bigg\}.
\end{align}
For $v_m\rightarrow q^{L_n}v_n$, writing
\begin{align}
    D_m&\equiv\det\left(\frac{1}{1-q^{L_i}v_i/v_j}\right)_{i,j=1,\dots,m}=\sum_{\sigma\in S_m}\sgn(\sigma)\prod_{i=1}^m\frac{1}{1-q^{L_i}v_i/v_{\sigma(i)}}\nonumber\\
    &=\sum_{\sigma\in S_m}\sgn(\sigma)\frac{v_m}{v_m-q^{L_{\sigma^{-1}(m)}}v_{\sigma^{-1}(m)}}\prod_{\substack{i=1\\\sigma(i)\neq m}}^m\frac{1}{1-q^{L_i}v_i/v_{\sigma(i)}},
\end{align}
we see that the contribution to the residue will come solely from the permutations $\sigma\in S_m$ for which $\sigma(n)=m$. This is due to the presence of the factor $(v_m-q^{L_n}v_n)$ in the numerator of the adjacent product. Therefore,
\begin{align}
    D_m&\rightarrow\frac{q^{L_n}v_n}{v_m-q^{L_n}v_{n}}\sum_{\substack{\sigma\in S_m\\ \sigma(n)=m}}\sgn(\sigma)\frac{1}{1-q^{L_m+L_n}v_n/v_{\sigma(m)}}\prod_{\substack{i=1\\i\neq n}}^{m-1}\frac{1}{1-q^{L_i}v_i/v_{\sigma(i)}}\nonumber\\
    &=-\frac{q^{L_n}v_n}{v_m-q^{L_n}v_n}\sum_{\tau\in S_{m-1}}\sgn(\tau)\prod_{i=1}^{m-1}\frac{1}{1-q^{L_i'}v_i/v_{\tau(i)}}\nonumber\\
    &=-\frac{q^{L_n}v_n}{v_m-q^{L_n}v_n}\det\left(\frac{1}{1-q^{L_i'}v_i/v_j}\right)_{i,j=1,\dots,m-1},
\end{align}
where $L_i'\equiv L_i+L_m\delta_{i,n}$. The second equality follows by writing the sum over permutations $\sigma\in S_m$ with $\sigma(n)=m$ as a sum over permutations $\tau\in S_{m-1}$ defined by $\tau(i)=\sigma(i)$ for $i\neq n$ and $\tau(n)=\sigma(m)$.\\\\
A straightforward calculation then reveals that
\begin{align}
    &\Res_{v_m\rightarrow q^{L_n}v_n}\Bigg\{\frac{\dd v_m}{2\pi iv_m}D_m^2\prod_{1\leq i<j\leq m}\frac{\left(v_j-q^{L_i}v_i\right)\left(v_i-q^{L_j}v_j\right)}{\left(v_j-v_i\right)\left(q^{L_i}v_i-q^{L_j}v_j\right)}\Bigg\}=-\frac{1-q^{L_n'}}{\left(1-q^{L_n}\right)\left(1-q^{L_m}\right)}\cdot\nonumber\\
    &\cdot\left[\det\left(\frac{1}{1-q^{L_i'}v_i/v_j}\right)_{i,j=1,\dots,m-1}\right]^2\prod_{1\leq i<j\leq m-1}\frac{\left(v_j-q^{L_i'}v_i\right)\left(v_i-q^{L_j'}v_j\right)}{\left(v_j-v_i\right)\left(q^{L_i'}v_i-q^{L_j'}v_j\right)},
\end{align}
with $L_i'=L_i+L_m\delta_{i,n}$.\\\\
Therefore, the residue at $v_m\rightarrow q^{L_n}v_n$ will be an expression analogous to equation (\ref{eqn:Fm}) in the remaining $m-1$ variables, with $L_i$ replaced by $L_i'$ and the overall factor becoming
\begin{equation}
    \left[\prod_{j=1}^m\left(1-q^{L_j}\right)\right]\rightarrow-\left[\prod_{j=1}^m\left(1-q^{L_j}\right)\right]\frac{1-q^{L_n'}}{\left(1-q^{L_n}\right)\left(1-q^{L_m}\right)}=-\left[\prod_{j=1}^{m-1}\left(1-q^{L_j'}\right)\right].
\end{equation}
Adding the residues from $v_m\rightarrow0$ and $v_m\rightarrow q^{L_n}v_n$, for all $n\in\{1,\dots,m-1\}$, we find the recurrence relation (\ref{eqn:RecurrenceHalfBPS}) listed in the introduction,
\begin{equation}
    F_{L_1,\dots,L_m}^{(m)}(q)=\frac{1}{1-q^{L_m}}F_{L_1,\dots,L_{m-1}}^{(m-1)}(q)-\sum_{n=1}^{m-1}F_{L_1',\dots,L_{m-1}'}^{(m-1)}(q),\label{eqn:314}
\end{equation}
with
\begin{equation}
    A_m(\zeta;q)=\frac{1}{m!}\sum_{k=m}^\infty(-1)^kq^{k(k+1)/2}\frac{\zeta q^k}{1-\zeta q^k}\sum_{\substack{L_1,\dots,L_m\geq1\\L_1+\dots+L_m=k}}F_{L_1,\dots,L_m}^{(m)}(q).\label{eqn:AmAns}
\end{equation}
Finally, for $m=1$, equation (\ref{eqn:Fm}) becomes
\begin{equation}
    F_{L_1}^{(1)}(q)=\left(1-q^{L_1}\right)\oint\frac{\dd v_1}{2\pi i v_1}\frac{1}{\left(1-q^{L_1}\right)^2}=\frac{1}{1-q^{L_1}}.
\end{equation}
\section{Results}\label{sec:results}
Having derived the equations listed in the introduction, we proceed to write expressions for the first few $G_N^{(m)}$, and check that the predictions for the giant-graviton contributions $\hat{Z}_k(q)$ extracted from the Fredholm determinant expansion match the known results. From equations (\ref{eqn:Amdef}) and (\ref{eqn:AmAns}), we find
\begin{equation}
    G_N^{(m)}(q)=\frac{1}{m!}\sum_{k=m}^\infty(-1)^kq^{Nk+k(k+1)/2}\sum_{\substack{L_1,\dots,L_m\geq1\\L_1+\dots+L_m=k}}F_{L_1,\dots,L_m}^{(m)}(q),
\end{equation}
which allows us to compute the terms in Murthy's Fredholm determinant expansion via the recurrence relation (\ref{eqn:RecurrenceHalfBPS}),
\begin{align}
    G_N^{(1)}(q)&=\sum_{k=1}^\infty(-1)^kq^{Nk+k(k+1)/2}\frac{1}{1-q^k}\nonumber\\
    &=-q^N\frac{q}{1-q}+q^{2N}\frac{q^3}{1-q^2}-q^{3N}\frac{q^6}{1-q^3}+q^{4N}\frac{q^{10}}{1-q^4}+\dots,\\
    G_N^{(2)}(q)&=\frac{1}{2}\sum_{k=2}^\infty(-1)^kq^{Nk+k(k+1)/2}\sum_{L_1=1}^{k-1}\left(\frac{1}{1-q^{L_1}}\frac{1}{1-q^{k-L_1}}-\frac{1}{1-q^k}\right)\nonumber\\
    &=q^{2N}\cdot\frac{q^3}{2}\left[\frac{1}{(1-q)^2}-\frac{1}{1-q^2}\right]-q^{3N}\cdot q^6\left[\frac{1}{(1-q)\left(1-q^2\right)}-\frac{1}{1-q^3}\right]+\nonumber\\
    &+q^{4N}\cdot\frac{q^{10}}{2}\left[\frac{2}{(1-q)\left(1-q^3\right)}+\frac{1}{\left(1-q^2\right)^2}-\frac{3}{1-q^4}\right]+\dots,\\
    G_N^{(3)}(q)&=-q^{3N}\cdot\frac{q^6}{6}\left[\frac{1}{(1-q)^3}-\frac{3}{(1-q)\left(1-q^2\right)}+\frac{2}{1-q^3}\right]+\nonumber\\
    &+q^{4N}\cdot\frac{q^{10}}{2}\left[\frac{1}{(1-q)^2\left(1-q^2\right)}-\frac{1}{\left(1-q^2\right)^2}-\frac{2}{(1-q)\left(1-q^3\right)}+\frac{2}{1-q^4}\right]+\dots,\\
    G_N^{(4)}(q)&=q^{4N}\cdot\frac{q^{10}}{24}\bigg[\frac{1}{(1-q)^4}-\frac{6}{(1-q)^2\left(1-q^2\right)}+\frac{8}{(1-q)\left(1-q^3\right)}+\nonumber\\
    &+\frac{3}{\left(1-q^2\right)^2}-\frac{6}{1-q^4}\bigg]+\dots.
\end{align}
One can immediately read off the giant-graviton contributions $\hat{Z}_k(q)$ from (\ref{eqn:ConnectingZk}) and (\ref{eqn:AmAns}),
\begin{equation}
    \hat{Z}_k(q)=(-1)^kq^{k(k+1)/2}\sum_{m=1}^k\frac{1}{m!}\sum_{\substack{L_1,\dots,L_m\geq1\\L_1+\dots+L_m=k}}F_{L_1,\dots,L_m}^{(m)}(q).\label{eqn:46}
\end{equation}
We can explicitly verify that the right side of the previous expression reproduces the known results (\ref{eqn:knownZkhat}),
\begin{align}
    \hat{Z}_1(q)&=-\frac{q}{1-q}=\frac{1}{1-q^{-1}},\\
    \hat{Z}_2(q)&=\frac{q^3}{(1-q)\left(1-q^2\right)}=\frac{1}{\left(1-q^{-1}\right)\left(1-q^{-2}\right)},\\
    \hat{Z}_3(q)&=-\frac{q^6}{(1-q)\left(1-q^2\right)\left(1-q^3\right)}=\frac{1}{\left(1-q^{-1}\right)\left(1-q^{-2}\right)\left(1-q^{-3}\right)},\\
    \hat{Z}_4(q)&=\frac{q^{10}}{(1-q)\left(1-q^2\right)\left(1-q^3\right)\left(1-q^4\right)}=\frac{1}{\left(1-q^{-1}\right)\left(1-q^{-2}\right)\left(1-q^{-3}\right)\left(1-q^{-4}\right)}.
\end{align}
In fact, starting from the recurrence relation (\ref{eqn:314}), it is possible to prove by a combinatorial argument \cite{Daniel} that
\begin{equation}
    \sum_{m=1}^k\frac{1}{m!}\sum_{\substack{L_1,\dots,L_m\geq1\\L_1+\dots+L_m=k}}F_{L_1,\dots,L_m}^{(m)}(q)=\prod_{n=1}^k\frac{1}{1-q^n},
\end{equation}
which allows us to recover the known result for the $1/2$-BPS giant-graviton contribution $\hat{Z}_k(q)$ from
equation (\ref{eqn:46}),
\begin{equation}
    \hat{Z}_k(q)=(-1)^kq^{k(k+1)/2}\prod_{n=1}^k\frac{1}{1-q^n}=\prod_{n=1}^k\frac{1}{1-q^{-n}}.
\end{equation}
We include the proof of \cite{Daniel} in appendix~\ref{app:B}.
\section{Superconformal Index as a Sum over Supermatrix Integrals}\label{sec:supermatrix}
We now come to the second main result of the paper, namely the identification of the terms in Murthy's Fredholm determinant expansion, $G_N^{(m)}(q)$, with the expectation value of the $N$th power of the Berezinian in a $U(m|m)$ superunitary matrix integral. The starting point is the algebraic version of the Fredholm determinant expansion, equation (\ref{eqn:startingpointforsupermatrix}),
\begin{align}
    \frac{Z_N\left(q\right)}{Z_\infty(q)}&=\sum_{m=0}^\infty\frac{(-1)^m}{\left(m!\right)^2}[1]\Bigg\{\left[\prod_{i=1}^m\left(\frac{u_i}{v_i}\right)^{N+1}\right]\cdot\left[\prod_{j=1}^mv_j^2\right]\cdot\left[\det\left(\frac{1}{v_j-u_i}\right)_{i,j=1,\dots,m}\right]^2\cdot\nonumber\\
    &\cdot\exp\left[\sum_{k=1}^\infty\frac{1}{k}\frac{f\left(q^k\right)}{\left(1-f\left(q^k\right)\right)}\sum_{i=1}^m\left(u_i^k-v_i^k\right)\sum_{j=1}^m\left(v_j^{-k}-u_j^{-k}\right)\right]\Bigg\},
\end{align}
This expression admits an analytic counterpart, where we work over the complex numbers, $u_j,v_i\in\mathbb{C}$, and replace the operation of extracting the constant coefficient in the power series in $u_j$ and $v_i$ with the corresponding contour integrals of the Laurent series. Concretely, the $m$th term in the Fredholm determinant expansion takes the form
\begin{align}
    G_N^{(m)}(q)&=\frac{(-1)^m}{\left(m!\right)^2}\prod_{j=1}^m\oint\frac{\dd u_j}{2\pi i}\prod_{i=1}^m\oint\frac{\dd v_i}{2\pi i }\Bigg\{\left[\prod_{i=1}^m\left(\frac{u_i}{v_i}\right)^N\right]\cdot\left[\det\left(\frac{1}{v_j-u_i}\right)_{i,j=1,\dots,m}\right]^2\cdot\nonumber\\
    &\cdot\exp\left[\sum_{k=1}^\infty\frac{1}{k}\frac{f\left(q^k\right)}{\left(1-f\left(q^k\right)\right)}\sum_{i=1}^m\left(u_i^k-v_i^k\right)\sum_{j=1}^m\left(v_j^{-k}-u_j^{-k}\right)\right]\Bigg\}.\label{eqn:5point2}
\end{align}
As mentioned previously, the contours are chosen to be circles of radius $r$ centered at the origin for all variables $u_j$, and circles of radius $R$ centered at the origin for all variables $v_i$, with $|q|<r/R<1$.\\\\
Employing the Cauchy determinant identity,
\begin{equation}
    \det\left(\frac{1}{v_j-u_i}\right)_{i,j=1,\dots,m}=\frac{\prod_{1\leq j<i\leq m}\left(u_j-u_i\right)\left(v_i-v_j\right)}{\prod_{i,j=1}^m\left(v_j-u_i\right)},
\end{equation}
one finds an expression reminiscent of the Vandermonde determinant of a physical supermatrix integral with ``positive-charge'' eigenvalues $u_j$ and ``negative-charge'' eigenvalues $v_i$. In fact, had the variables $u_j$ and $v_i$ been unit-norm (for example by taking $r=1/R<1$ and then taking $R\rightarrow1^+$), we would have found
\begin{equation}
    \left[\frac{\prod_{1\leq j<i\leq m}\left(u_j-u_i\right)\left(v_i-v_j\right)}{\prod_{i,j=1}^m\left(v_j-u_i\right)}\right]^2=(-1)^{m^2}\left(\prod_{i=1}^m\frac{1}{u_iv_i}\right)\frac{\prod_{1\leq j<i\leq m}\left|u_i-u_j\right|^2\left|v_i-v_j\right|^2}{\prod_{i,j=1}^m\left|u_i-v_j\right|^2}
\end{equation}
Plugging this into (\ref{eqn:5point2}), we find
\begin{align}
    G_N^{(m)}(q)&=\frac{1}{m!}\prod_{j=1}^m\oint\frac{\dd u_j}{2\pi iu_j}\cdot\frac{1}{m!}\prod_{i=1}^m\oint\frac{\dd v_i}{2\pi iv_i }\Bigg\{\left[\prod_{i=1}^m\left(\frac{u_i}{v_i}\right)^N\right]\frac{\prod_{1\leq j<i\leq m}\left|u_i-u_j\right|^2\left|v_i-v_j\right|^2}{\prod_{i,j=1}^m\left|u_i-v_j\right|^2}\nonumber\\
    &\cdot\exp\left[\sum_{k=1}^\infty\frac{1}{k}\frac{f\left(q^k\right)}{\left(1-f\left(q^k\right)\right)}\sum_{i=1}^m\left(u_i^k-v_i^k\right)\sum_{j=1}^m\left(v_j^{-k}-u_j^{-k}\right)\right]\Bigg\}.
\end{align}
Identifying the variables $u_j$ and $v_i$ as the eigenvalues $e^{i\theta_j}$, $e^{i\phi_i}$ of a $U(m|m)$ unitary supermatrix $W$, we have the following expressions for the superdeterminant, the supertraces, and the Vandermonde determinant:
\begin{align}
    \Ber(W)&=\prod_{i=1}^m\left(\frac{u_i}{v_i}\right),\\
    \Str\left(W^k\right)&=\sum_{i=1}^m\left(u_i^k-v_i^k\right),\\
    \Str\left(W^{-k}\right)&=\sum_{j=1}^m\left(u_j^{-k}-v_j^{-k}\right),\\
    \VdM(W)&=\frac{\prod_{1\leq j<i\leq m}\left|u_i-u_j\right|^2\left|v_i-v_j\right|^2}{\prod_{i,j=1}^m\left|u_i-v_j\right|^2}.
\end{align}
Thus, we arrive at the supermatrix integral expression for the $m$th term in the Fredholm determinant expansion of the superconformal index:
\begin{equation}
    G_N^{(m)}(q)=\int_{U(m|m)}dW\;\Ber(W)^N\exp\left[-\sum_{k=1}^\infty\frac{1}{k}\frac{f\left(q^k\right)}{\left(1-f\left(q^k\right)\right)}\Str\left(W^k\right)\Str\left(W^{-k}\right)\right].
\end{equation}
\section{Summary and Discussion}\label{sec:conclusions}
Having completed the mathematical derivation of our two results, we now summarize and discuss their potential implications.\\\\
Our first result consists of a prescription for extracting the terms of the giant-graviton expansion of the superconformal index,
\begin{equation}
    \frac{Z_N(q)}{Z_\infty(q)}=1+\sum_{k=1}^\infty q^{kN}\hat{Z}_k(q),
\end{equation}
from Murthy's Fredholm determinant expansion,
\begin{equation}
    \frac{Z_N(q)}{Z_\infty(q)}=1+\sum_{m=1}^\infty G_N^{(m)}(q).
\end{equation}
We carried out this algorithm for the particularly simple case of the $1/2$-BPS index, where we showed that the terms of the second expansion take the form
\begin{equation}
    G_N^{(m)}(q)=\frac{1}{m!}\sum_{k=m}^\infty(-1)^kq^{Nk+k(k+1)/2}\sum_{\substack{L_1,\dots,L_m\geq1\\L_1+\dots+L_m=k}}F_{L_1,\dots,L_m}^{(m)}(q),
\end{equation}
where the terms $F_{L_1,\dots,L_m}^{(m)}(q)$ satisfy the following recurrence relation for $m\geq2$,
\begin{equation}
    F_{L_1,\dots,L_m}^{(m)}(q)=\frac{1}{1-q^{L_m}}F_{L_1,\dots,L_{m-1}}^{(m-1)}(q)-\sum_{j=1}^{m-1}F_{L_1,\dots,L_{j-1},L_j+L_m,L_{j+1},\dots,L_{m-1}}^{(m-1)}(q),
\end{equation}
with
\begin{equation}
    F_L^{(1)}(q)=\frac{1}{1-q^L}.
\end{equation}
Our prescription then led to the prediction for the giant-graviton contributions $\hat{Z}_k(q)$,
\begin{equation}
    \hat{Z}_k(q)=(-1)^kq^{k(k+1)/2}\sum_{m=1}^k\frac{1}{m!}\sum_{\substack{L_1,\dots,L_m\geq1\\L_1+\dots+L_m=k}}F_{L_1,\dots,L_m}^{(m)}(q).
\end{equation}
One can further prove \cite{Daniel} this simplifies to the expected expression for the associated $U(k)$ gauge theory index,
\begin{equation}
    \hat{Z}_k(q)=(-1)^kq^{k(k+1)/2}\prod_{n=1}^k\frac{1}{1-q^n}.
\end{equation}
Of course, the final answer for the $1/2$-BPS case is well-known. Our choice to analyze this example is obviously not motivated by a hope to learn more about this particular case, but instead by its potential for generalization to other examples where the final answer might not be known. The goal of our analysis is to provide a proof of concept that the prescription presented here allows us to extract the giant-graviton contributions $\hat{Z}_k(q)$ from the Fredholm determinant expansion of the index, with the simplicity of the $1/2$-BPS case making it the ideal candidate for our case study.\\\\
We are not ready to conclude that the current analysis provides a proof of existence for the giant-graviton expansion of general indices.\footnote{As in the rest of the paper, by giant-graviton expansion we specifically mean an expansion of the index of the form $Z_N(q)/Z_\infty(q)=1+\sum_{k=1}^\infty q^{kN}\hat{Z}_k(q)$, with $\hat{Z}_k(q)$ independent of $N$.} There are two reasons for this which we want to highlight. The first is simply that we have not been precise about the conditions on the coefficients $\hat{a}_l$ in the expansion of (\ref{eqn:defahatl}),
\begin{equation}
    \frac{f(q)}{1-f(q)}=\sum_{l=1}^\infty\hat{a}_lq^l,\;\;\;\hat{a}_l\in\mathbb{Z},\nonumber
\end{equation}
under which equation (\ref{eqn:210}) holds,
\begin{align}
    \frac{\mathcal{Z}(\zeta;q)}{Z_\infty(q)}&=\frac{\zeta}{1-\zeta}+\sum_{m=1}^\infty\frac{(-1)^m}{\left(m!\right)^2}\prod_{j=1}^m\oint\frac{\dd u_j}{2\pi i}\prod_{i=1}^m\oint\frac{\dd v_i}{2\pi i v_i^2}\Bigg\{\frac{\zeta\prod_{i=1}^m\left(u_i/v_i\right)}{1-\zeta\prod_{i=1}^m\left(u_i/v_i\right)}\cdot\nonumber\\&\cdot\left[\det\left(\frac{1}{1-u_i/v_j}\right)_{i,j=1,\dots,m}\right]^2\prod_{i,j=1}^m\prod_{l=1}^\infty\left[\frac{\left(1-q^lu_i/u_j\right)\left(1-q^lv_i/v_j\right)}{\left(1-q^lu_i/v_j\right)\left(1-q^lv_i/u_j\right)}\right]^{\hat{a}_l}\Bigg\}.\nonumber
\end{align}
However, we expect to be able to rigorously prove that the residue prescription we have presented is valid under such a set of conditions, and we plan to clarify this point in upcoming work. The other point we want to highlight is that in cases where some of the coefficients $\hat{a}_l$ are larger or equal to 2 in absolute value, leading to higher-order poles in the integrand of equation (\ref{eqn:210}), calculating the residues associated to those poles involves taking a number of derivatives of the rest of the expression in the integrand. This follows from a standard application of the residue theorem for a holomorphic function $f(z)$,
\begin{equation}
    \oint\frac{\dd z}{2\pi i}\frac{f(z)}{(z-w)^n}=\frac{f^{(n-1)}(w)}{(n-1)!}.\nonumber
\end{equation}
A number of these derivatives can act on the $\zeta$-dependent factor in the integrand, subsequently producing higher-order poles in $\zeta$ in the final answer. However, the existence of a giant-graviton expansion for the index $Z_N(q)$ requires that the final answer for $\mathcal{Z}(\zeta;q)/\mathcal{Z}_\infty(q)$ be expressible as a sum of simple poles in $\zeta$,
\begin{equation}
    \frac{\mathcal{Z}(\zeta;q)}{Z_\infty(q)}=\frac{\zeta}{1-\zeta}+\sum_{k=1}^\infty\frac{\zeta q^k}{1-\zeta q^k}\hat{Z}_k(q).\nonumber
\end{equation}
The appearance of higher-order poles in $\zeta$ derived from our algorithm is then an indicator that the index whose generating function we are computing should not have a giant-graviton expansion, unless cancellations of these higher-order poles eventually occur. Having not yet analyzed under what circumstances such cancellations could occur, we cannot determine the general conditions under which an index admits a giant-graviton expansion. However, given a particular index, one can use our procedure to determine its giant-graviton expansion, provided the expansion exists.\\\\
One might argue that if we already expect a giant-graviton expansion to exist for some index $Z_N(q)$, given as a power series in $q$, one could just organize all the indices $Z_N(q)$ into the generating function $\mathcal{Z}(\zeta;q)$, and read off the giant-graviton contributions from the poles of the analytic continuation of $\mathcal{Z}(\zeta;q)$ directly. This is true. However, in general, the analytic continuation of $\mathcal{Z}(\zeta;q)$ might not be easy to determine from its form as a power series in $q$ and $\zeta$. The advantage of the procedure we described is that the $\zeta$-dependence of $\mathcal{Z}(\zeta;q)$ appearing in equation (\ref{eqn:210}) is already in the form of a sum of simple poles in $\zeta$. As one performs the integrals in the variables $u_j$ and $v_i$, the resulting answer organizes into a sum of poles in $\zeta$ from which one can read off the giant-graviton contributions. It would be very interesting to generalize the present discussion to the case of indices depending on multiple fugacities such as the $1/16$-BPS index, which we hope to address in future work.\\\\
Our second result consists of the identification of the terms in Murthy's Fredholm determinant expansion of the $U(N)$ index,
\begin{equation}
    \frac{Z_N(q)}{Z_\infty(q)}=1+\sum_{m=1}^\infty G_N^{(m)}(q),
\end{equation}
with the expectation value of the $N$th power of the Berezinian in the $U(m|m)$ superunitary matrix integral,
\begin{equation}
    G_N^{(m)}(q)=\int_{U(m|m)}\dd W\;\Ber(W)^N\exp\left[-\sum_{k=1}^\infty\frac{1}{k}\frac{f\left(q^k\right)}{\left(1-f\left(q^k\right)\right)}\Str\left(W^k\right)\Str\left(W^{-k}\right)\right],
\end{equation}
whose \textit{definition}\footnote{The fact that the supermatrix integral is defined by the corresponding integrals over eigenvalues is also true in the case of the $U(n|m)$-symmetric Hermitian supermatrix integral \cite{SchiappaEigenvalues}.} is given by equation (\ref{eqn:5point2}),
\begin{align}
    G_N^{(m)}(q)&=\frac{(-1)^m}{\left(m!\right)^2}\prod_{j=1}^m\oint\frac{\dd u_j}{2\pi i}\prod_{i=1}^m\oint\frac{\dd v_i}{2\pi i }\Bigg\{\left[\prod_{i=1}^m\left(\frac{u_i}{v_i}\right)^N\right]\cdot\left[\det\left(\frac{1}{v_j-u_i}\right)_{i,j=1,\dots,m}\right]^2\cdot\nonumber\\
    &\cdot\exp\left[-\sum_{k=1}^\infty\frac{1}{k}\frac{f\left(q^k\right)}{\left(1-f\left(q^k\right)\right)}\sum_{i=1}^m\left(u_i^k-v_i^k\right)\sum_{j=1}^m\left(u_j^{-k}-v_j^{-k}\right)\right]\Bigg\},
\end{align}
with contours chosen to be circles of radius $r$ centered at the origin for all variables $u_j$, and circles of radius $R$ centered at the origin for all variables $v_i$, with $|q|<r/R<1$.\\\\
Hermitian analogs of this type of supermatrix integrals (called ``physical supermatrix integrals'') were the main object of study in a remarkable recent work by Mariño, Schiappa, and Schwick \cite{SchiappaEigenvalues} in the context of nonperturbative effects in minimal string theory \cite{MinimalStrings1, MinimalStrings2}, a type of Liouville quantum gravity. It is well-known that the $(2,p)$ minimal string theory enjoys a dual description in terms of a double-scaled Hermitian matrix integral \cite{QG1, QG2, QG3, QG4, QG5}. In this description, perturbative fluctuations around the one-cut saddle of the matrix integral reproduce the contributions associated to perturbative closed string diagrams. Additionally, fluctuations around instanton-saddles of the matrix integral in which a number $m$ of eigenvalues have tunneled away from the cut, to an extremum of the effective potential, correspond to string diagrams allowed to end on a stack of $m$ ZZ branes \cite{ZZ}. The simplicity of $(2,p)$ minimal string theory allowed a systematic analysis of these phenomena and of the resurgence properties of the perturbation series associated to these models \cite{SchiappaInstanton1, SchiappaInstanton2, SchiappaInstanton3, SchiappaInstanton4, SchiappaInstanton5}.\\\\
It was realized \cite{Marino1, SchiappaInstanton3} that in addition to the standard ZZ branes of Liouville theory, resurgence required the existence of additional objects which seemed to behave as ``ZZ branes with opposite tension'' (sometimes called ghost branes or negative branes) on the string theory side and as ``opposite-charge eigenvalues'' (sometimes called anti-eigenvalues) on the matrix integral side. The study by Mariño, Schiappa, and Schwick \cite{SchiappaEigenvalues} established that the right language in which to describe these objects, on the matrix integral side, is that of physical Hermitian supermatrices. In this framework, the standard $U(N)$-symmetric Hermitian matrix being integrated is upgraded to a $U(N|M)$-symmetric physical Hermitian supermatrix, in which one is allowed to study the tunnelling of both a subset of the $N$ eigenvalues and a subset of the $M$ anti-eigenvalues. It was shown that this description perfectly reproduces the predictions obtained from resurgence. Subsequently, Schiappa, Schwick, and Tamarin \cite{SchiappaBranes} established through detailed Liouville BCFT calculations the existence of the opposite-tension ZZ branes, and showed that the contributions from these objects matched the matrix integral computations.\\\\
One might speculate that, similarly, the $U(m|m)$ superunitary matrix integral identification of the $m$th term $G_N^{(m)}(q)$ of Murthy's Fredholm determinant expansion of the superconformal index hints at a $U(m|m)$ gauge theory origin for the term, with the gauge theory perhaps appearing as the worldvolume theory of a stack of $m$ branes and $m$ negative branes. Supergroup gauge theories realized as worldvolume theories of brane/negative-brane stacks have been studied in the past by Dijkgraaf, Heidenreich, Jefferson, and Vafa \cite{DHJV}.\footnote{For a recent survey of supergroup gauge theory, see also \cite{supergroupsurvey} and references therein.} If such a supergroup gauge theory origin of the term $G_N^{(m)}(q)$ were to exist, it would complement the identification of the $k$th term in the Gaiotto-Lee expansion with the index of an associated $U(k)$ gauge theory describing the worldvolume theory of a stack of $k$ giant-graviton branes.\\\\
However, one might also note that the identification of the $m$th term in the Fredholm determinant expansion with a $U(m|m)$ supermatrix integral holds for Fredholm determinants (or Toeplitz determinants that can be written as Fredholm determinants \cite{BO,GC}) which are not related to superconformal indices. It would be interesting to understand the implications of the supermatrix integral presentation in the context of other Fredholm (or Toeplitz) determinants.\\\\
Finally, it would be interesting to study the possibility of a connection between the terms in these expansions and eigenvalue instantons in the associated matrix integrals. We hope to address this goal in future work.
\appendix
\section{Fredholm Determinant Expansion} \label{app:A}
In this appendix, we review the derivation of the expansion of the superconformal index proposed in \cite{Murthy}. We also review the evaluation of Fredholm determinants which was derived by Tracy and Widom \cite{TW1, TW2} in the context of the 2D Ising model. This evaluation was also considered by Liu and Rajappa \cite{LiuRajappa} in the present context. Concretely, we review the derivation of the expansion of matrix integrals of two possible forms:
\begin{align}
    \Tilde{Z}_N\left(t_k^+,t_k^-\right)&\equiv\int_{U(N)}\dd U\;\exp\left[\sum_{k=1}^\infty\left(\frac{t_k^+}{k}\Tr\left(U^k\right)+\frac{t_k^-}{k}\Tr{\left(U^{-k}\right)}\right)\right], \label{Z_NTilde}\\
    Z_N\left(g_k\right)&\equiv\int_{U(N)}\dd U\;\exp\left[\sum_{k=1}^\infty\frac{g_k}{k}\Tr\left(U^k\right)\Tr\left(U^{-k}\right)\right], \label{Z_N}
\end{align}
where the integrals should be interpreted as formal power series in variables $t_k^\pm$ and $g_k$ with $k\in\mathbb{N}^*$, with coefficients in $\mathbb{C}$.
The integral (\ref{Z_N}) can be obtained from (\ref{Z_NTilde}), via the Hubbard-Stratonovich transform \cite{Hubbard, Strat} which, in this context, is simply the map between power series defined by
\begin{equation}
    \left(t_k^+\right)^{n}\left(t_k^-\right)^{n'}\rightarrow n!\;k^ng_k^n\;\delta_{n,n'}.\label{eqn:SimpleHS}
\end{equation}
One can show that the extension of this map applied to (\ref{Z_NTilde}) for all $k\in\mathbb{N}^*$ leads to (\ref{Z_N}),
\begin{equation}
    \Tilde{Z}_N\left(t_k^+,t_k^-\right)\rightarrow Z_N\left(g_k\right).
\end{equation}
The Hubbard-Stratonovich transform can be represented formally by a Gaussian integral in all pairs of variables $\left(t_k^+,t_k^-\right)$, where one thinks of $t_k^+$ and $t_k^-$ as complex conjugates of each other, with the integral running over the entire complex plane:
\begin{equation}
    Z_N\left(g_k\right)=\prod_{n=1}^\infty\int_{\mathbb{C}}\frac{dt_n^+dt_n^-}{2\pi n g_n}e^{-t_n^+t_n^-/\left(ng_n\right)}\;\Tilde{Z}_N\left(t_k^+,t_k^-\right). \label{HSZZ}
\end{equation}
However, we stress that in our context, (\ref{HSZZ}) should only be interpreted as a notational convention for the Hubbard-Stratonovich transform which is properly defined as a map between formal power series.\\\\
A key feature of matrix integrals with $U(N)$ symmetry is that one can gauge fix and reduce the integral to one solely over the eigenvalues of the matrix. In the case of (\ref{Z_NTilde}), letting $z_l$ denote the eigenvalues of the unitary matrix $U$, one finds
\begin{equation}
    \Tilde{Z}_N\left(t_k^+,t_k^-\right)=\frac{1}{N!}\prod_{l=1}^N\oint\frac{\dd z_l}{2\pi i z_l}\prod_{1\leq i<j\leq N}\left|z_i-z_j\right|^2\prod_{i=1}^N\exp\left[\sum_{k=1}^\infty\left(\frac{t_k^+}{k}z_i^k+\frac{t_k^-}{k}z_i^{-k}\right)\right].\label{eqn:ZNTilde_Diagonal}
\end{equation}
The first step in deriving the Fredholm determinant expansion of the index consists of writing the unitary matrix integral $\Tilde{Z}_N\left(t_k^+,t_k^-\right)$ as the determinant of a Toeplitz matrix.
\subsection{Unitary matrix integral as a Toeplitz determinant}
Let
\begin{equation}
    \varphi(\xi)\equiv\exp\left[\sum_{k=1}^\infty\left(\frac{t_k^+}{k}\xi^k+\frac{t_k^-}{k}\xi^{-k}\right)\right]
\end{equation}
be a symbol defined in a neighborhood of the unit circle $|\xi|=1$. $\varphi(\xi)$ should technically be thought of as a compact expression for a $\xi$-dependent power series in $t_k^+$, $t_k^-$. Letting $\varphi_j$ denote the $j$th Fourier coefficient of $\varphi$ on the unit circle,
\begin{equation}
    \varphi_j\equiv\frac{1}{2\pi i}\oint\dd \xi\;\varphi(\xi)\xi^{-(j+1)},
\end{equation}
and organizing the Fourier coefficients into an $N\times N$ Toeplitz matrix, whose entries are given by $T_{ij}\equiv\varphi_{i-j}$, one finds that the determinant of this matrix agrees with the unitary matrix integral (\ref{eqn:ZNTilde_Diagonal}):
\begin{equation}
    \det(T)=\Tilde{Z}_N\left(t_k^+,t_k^-\right).
\end{equation}
This follows from the standard identity \cite{Andreief},
\begin{equation}
    \det\left(\int d\mu(z)\;f_j(z)g_k(z)\right)=\frac{1}{N!}\int d\mu\left(z_1\right)\dots\int d\mu\left(z_N\right)\left[\det\left(f_j\left(z_k\right)\right)\det\left(g_j\left(z_k\right)\right)\right],
\end{equation}
with
\begin{equation}
    f_j(z)=z^{1-j},\;\;\;g_k(z)=z^{k-1},\;\;\;d\mu(z)=\frac{\dd z}{2\pi i z}\varphi(z),
\end{equation}
and the expression for the determinant of a Vandermonde matrix with entries $V_{i,j}=z_i^{j-1}$,
\begin{equation}
    \det(V)=\prod_{1\leq i<j\leq N}\left(z_j-z_i\right).
\end{equation}
Having reviewed the identification of the unitary matrix integral with a Toeplitz determinant, we proceed by restating the celebrated theorem of Borodin, Okounkov, Geronimo, and Case \cite{BO,GC}. This will provide an expression for $\Tilde{Z}_N\left(t_k^+,t_k^-\right)$ as an associated Fredholm determinant, which takes the form of a convergent series.
\subsection{Toeplitz determinant as a Fredholm determinant}
We will not reproduce the proof of the BOGC theorem, but merely recall the result. We are interested in the algebraic version of the theorem. Consider the symbol
\begin{equation}
    V^*(\zeta)\equiv\sum_{k=1}^\infty\left[\frac{t_k^-}{k}\zeta^{-k}-\frac{t_k^+}{k}\zeta^k\right],
\end{equation}
and the kernel $\kappa(i,j)$ defined as the coefficient of $\left(\zeta^i,\eta^{-j}\right)$ in the generating function
\begin{equation}
    \sum_{i,j\in\mathbb{Z}}u^{-i}v^j\kappa(i,j)=\exp\left[V^*(1/u)-V^*(1/v)\right]\sum_{l=1}^\infty\left(\frac{u}{v}\right)^l.
\end{equation}
Letting
\begin{equation}
    \Tilde{Z}_\infty\left(t_k^+,t_k^-\right)\equiv\exp\left(\sum_{k=1}^\infty\frac{t_k^+t_k^-}{k}\right),
\end{equation}
the BOGC theorem states that
\begin{equation}
    \det(T)=\Tilde{Z}_\infty\left(t_k^+,t_k^-\right)\sum_{m=0}^\infty(-1)^m\sum_{N\leq l_1<l_2<\dots<l_m}^\infty\det\left[\kappa(l_i,l_j)\right]_{i,j=1,\dots,m}.
\end{equation}
Therefore,
\begin{equation}
    \frac{\Tilde{Z}_N\left(t_k^+,t_k^-\right)}{\Tilde{Z}_\infty\left(t_k^+,t_k^-\right)}=\sum_{m=0}^\infty(-1)^m\sum_{N\leq l_1<l_2<\dots<l_m}^\infty\det\left[\kappa(l_i,l_j)\right]_{i,j=1,\dots,m}.
\end{equation}
Note that if $l_i=l_j$, then
\begin{equation}
    \det\left[\kappa\left(l_p,l_q\right)\right]_{p,q=1,\dots,m}=0 
\end{equation}
 since the $i$th and $j$th row of the matrix will be identical. Additionally, swapping $l_i$ and $l_j$ will correspond to swapping both the $i$th and $j$th rows of the matrix and the $i$th and $j$th columns of the matrix. Therefore, the determinant does not change under swapping $l_i$ and $l_j$. We can thus write,
 \begin{equation}
    \frac{\Tilde{Z}_N\left(t_k^+,t_k^-\right)}{\Tilde{Z}_\infty\left(t_k^+,t_k^-\right)}=\sum_{m=0}^\infty\frac{(-1)^m}{m!}\sum_{N\leq l_1,l_2,\dots,l_m}^\infty\det\left[\kappa(l_i,l_j)\right]_{i,j=1,\dots,m}.
\end{equation}
Letting
\begin{equation}
    \kappa_N\left(p_i,p_j\right)=\kappa\left(N+p_i,N+p_j\right),
\end{equation}
we have
\begin{equation}
    \frac{\Tilde{Z}_N\left(t_k^+,t_k^-\right)}{\Tilde{Z}_\infty\left(t_k^+,t_k^-\right)}=\sum_{m=0}^\infty\frac{(-1)^m}{m!}\sum_{0\leq p_1,p_2,\dots,p_m}\det\left[\kappa_N\left(p_i,p_j\right)\right]_{i,j=1,\dots,m},\label{eqn:Fred}
\end{equation}
with
\begin{align}
    \sum_{i,j\in\mathbb{Z}}u^{-i-N}v^{j+N}\kappa_N(i,j)&=\exp\left[V^*\left(u^{-1}\right)-V^*\left(v^{-1}\right)\right]\sum_{l=1}^\infty\left(\frac{u}{v}\right)^l\nonumber\\
    &=\exp\left[\sum_{k=1}^\infty\left(\frac{t_k^-}{k}\left(u^k-v^k\right)-\frac{t_k^+}{k}\left(u^{-k}-v^{-k}\right)\right)\right]\sum_{l=1}^\infty\left(\frac{u}{v}\right)^l.
\end{align}
Therefore, we can write
\begin{equation}
    \kappa_N(p_i,p_j)=[1]\left\{u^{p_i+N}v^{-p_j-N}\exp\left[\sum_{k=1}^\infty\left(\frac{t_k^-}{k}\left(u^k-v^k\right)-\frac{t_k^+}{k}\left(u^{-k}-v^{-k}\right)\right)\right]\sum_{l=1}^\infty\left(\frac{u}{v}\right)^l\right\},
\end{equation}
where the notation $[1]\{\dots\}$ denotes extracting the constant term from the power series in $u$ and $v$ within the brackets.
\subsection{Tracy-Widom evaluation of a Fredholm determinant}
The remaining issue is to evaluate the sum over $p_j$ on the right side of (\ref{eqn:Fred}). It turns out that this calculation has been considered in the literature before, albeit in a different context, that of studying the susceptibility of the 2D Ising model \cite{TW1,TW2}. The same calculation was also performed in \cite{LiuRajappa} in the context of the giant-graviton expansion. We reproduce the main ideas here.\\\\
First, let
\begin{equation}
    F(u,v)=\frac{1}{1-u/v}u^{p_i+N+1}v^{-p_j-N-1}\exp\left[\sum_{k=1}^\infty\left(\frac{t_k^-}{k}\left(u^k-v^k\right)-\frac{t_k^+}{k}\left(u^{-k}-v^{-k}\right)\right)\right],
\end{equation}
where the fraction $1/(1-u/v)$ is understood as the power series in $u/v$ obtained by expanding for $|u|<|v|$. We can write the summand in (\ref{eqn:Fred}) as
\begin{align}
    \det\left[\kappa_N\left(p_i,p_j\right)\right]_{i,j=1,\dots,m}&=\varepsilon_{s_1,\dots,s_m}\kappa_N\left(p_1,p_{s_1}\right)\dots\kappa_N\left(p_m,p_{s_m}\right)\nonumber\\
    &=\varepsilon_{s_1,\dots,s_m}\Big([1]F\left(u_1,v_{s_1}\right)\Big)\dots\Big([1]F\left(u_m,v_{s_m}\right)\Big).
\end{align}
Since each $u_i$ and each $v_j$ only appear in one of the $m$ factors on the right side of the previous expression, we find
\begin{align}
    \det\left[\kappa_N\left(p_i,p_j\right)\right]_{i,j=1,\dots,m}&=[1]\Big\{\varepsilon_{s_1,\dots,s_m}F\left(u_1,v_{s_1}\right)\dots F\left(u_m,v_{s_m}\right)\Big\}\nonumber\\
    &=[1]\Bigg\{\varepsilon_{s_1,\dots,s_m}\prod_{i=1}^m\left(\frac{u_i}{v_i}\right)^{N+1}\prod_{i=1}^m\left(\frac{u_i}{v_i}\right)^{p_i}\left(\prod_{j=1}^m\frac{1}{1-u_j/v_{s_j}}\right)\cdot\nonumber\\
    &\cdot\exp\left[\sum_{k=1}^\infty\left(\frac{t_k^-}{k}\sum_{i=1}^m\left(u_i^k-v_i^k\right)+\frac{t_k^+}{k}\sum_{i=1}^m\left(v_i^{-k}-u_i^{-k}\right)\right)\right]\Bigg\}.
\end{align}
Therefore, $\det\left[\kappa_N\left(p_i,p_j\right)\right]_{i,j=1,\dots,m}$ can be identified with the constant coefficient of a power series in $2m$ variables:
\begin{align}
    \det\left[\kappa_N\left(p_i,p_j\right)\right]_{i,j=1,\dots,m}&=[1]\Bigg\{\prod_{i=1}^m\left(\frac{u_i}{v_i}\right)^{N+1+p_i}\cdot\det\left(\frac{1}{1-u_i/v_j}\right)_{i,j=1,\dots,m}\cdot\nonumber\\
    &\cdot\exp\left[\sum_{k=1}^\infty\left(\frac{t_k^-}{k}\sum_{i=1}^m\left(u_i^k-v_i^k\right)+\frac{t_k^+}{k}\sum_{i=1}^m\left(v_i^{-k}-u_i^{-k}\right)\right)\right]\Bigg\}.
\end{align}
Summing over the $p_i$,
\begin{align}
    \sum_{0\leq p_1,\dots,p_m}&\det\left[\kappa_N\left(p_i,p_j\right)\right]_{i,j=1,\dots,m}=[1]\Bigg\{\prod_{i=1}^m\left(\frac{u_i}{v_i}\right)^{N+1}\cdot\det\left(\frac{1}{1-u_i/v_j}\right)_{i,j=1,\dots,m}\cdot\nonumber\\
    &\cdot\prod_{i=1}^m\frac{1}{1-u_i/v_i}\cdot\exp\left[\sum_{k=1}^\infty\left(\frac{t_k^-}{k}\sum_{i=1}^m\left(u_i^k-v_i^k\right)+\frac{t_k^+}{k}\sum_{i=1}^m\left(v_i^{-k}-u_i^{-k}\right)\right)\right]\Bigg\}. \label{eqn:sumkappaN}
\end{align}
Since we're only interested in the constant term of the power series in the previous expression, we can make the replacement
\begin{equation}
    \det\left(\frac{1}{1-u_i/v_j}\right)_{i,j=1,\dots,m}\prod_{i=1}^m\frac{1}{1-u_i/v_i}\rightarrow\frac{1}{m!}\det\left(\frac{1}{1-u_i/v_j}\right)_{i,j=1,\dots,m}^2.
\end{equation}
This can be seen as follows: Writing
\begin{equation}
    \det\left(\frac{1}{1-u_i/v_j}\right)_{i,j=1,\dots,m}\prod_{i=1}^m\frac{1}{1-u_i/v_i}=\sum_{\tau\in S_m}\sgn(\tau)\prod_{j=1}^m\frac{1}{1-u_j/v_{\tau(j)}}\cdot\prod_{i=1}^m\frac{1}{1-u_i/v_i},
\end{equation}
for a fixed permutation $\sigma\in S_m$, we can perform the relabeling $v_i\rightarrow v_{\sigma(i)}$, which will not affect any other part of the expression within the $\{\dots\}$, since the other parts are symmetric under the relabeling. This relabeling corresponds to
\begin{align}
    \det\left(\frac{1}{1-u_i/v_j}\right)_{i,j=1,\dots,m}&\prod_{i=1}^m\frac{1}{1-u_i/v_i}\rightarrow\sum_{\tau\in S_m}\sgn(\tau)\prod_{j=1}^m\frac{1}{1-u_j/v_{(\sigma\circ\tau)(j)}}\cdot\prod_{i=1}^m\frac{1}{1-u_i/v_{\sigma(i)}}\nonumber\\
    &=\left[\sum_{\rho\in S_m}\sgn(\rho)\prod_{j=1}^m\frac{1}{1-u_j/v_{\rho(j)}}\right]\cdot\left[\sgn(\sigma)\prod_{i=1}^m\frac{1}{1-u_i/v_{\sigma(i)}}\right].
\end{align}
and does not affect the constant coefficient of the power series within the brackets $\{\dots\}$. Therefore, summing over all $\sigma\in S_m$, leads to a power series whose constant coefficient is $m!$ times the constant coefficient of the original power series, and to two copies of the determinant on the right side of the previous expression. Therefore, one can replace
\begin{align}
    \det\left(\frac{1}{1-u_i/v_j}\right)_{i,j=1,\dots,m}&\prod_{i=1}^m\frac{1}{1-u_i/v_i}\rightarrow\frac{1}{m!}\det\left(\frac{1}{1-u_i/v_j}\right)_{i,j=1,\dots,m}^2.
\end{align}
The expression (\ref{eqn:sumkappaN}) becomes
\begin{align}
    \sum_{0\leq p_1,\dots,p_m}&\det\left[\kappa_N\left(p_i,p_j\right)\right]_{i,j=1,\dots,m}=[1]\Bigg\{\prod_{i=1}^m\left(\frac{u_i}{v_i}\right)^{N+1}\cdot\frac{1}{m!}\det\left(\frac{1}{1-u_i/v_j}\right)_{i,j=1,\dots,m}^2\cdot\nonumber\\
    &\cdot\exp\left[\sum_{k=1}^\infty\left(\frac{t_k^-}{k}\sum_{i=1}^m\left(u_i^k-v_i^k\right)+\frac{t_k^+}{k}\sum_{i=1}^m\left(v_i^{-k}-u_i^{-k}\right)\right)\right]\Bigg\}.
\end{align}
\subsection{Fredholm determinant expansion of the index}
Putting the previous expression together with (\ref{eqn:Fred}), we find
\begin{align}
    \Tilde{Z}_N\left(t_k^+,t_k^-\right)&=\Tilde{Z}_\infty\left(t_k^+,t_k^-\right)\sum_{m=0}^\infty\frac{(-1)^m}{\left(m!\right)^2}[1]\Bigg\{\prod_{i=1}^m\left(\frac{u_i}{v_i}\right)^{N+1}\cdot\det\left(\frac{1}{1-u_i/v_j}\right)_{i,j=1,\dots,m}^2\cdot\nonumber\\
    &\cdot\exp\left[\sum_{k=1}^\infty\left(\frac{t_k^-}{k}\sum_{i=1}^m\left(u_i^k-v_i^k\right)+\frac{t_k^+}{k}\sum_{i=1}^m\left(v_i^{-k}-u_i^{-k}\right)\right)\right]\Bigg\}.
\end{align}
Finally, one can perform the Hubbard-Stratonovich transform (\ref{eqn:SimpleHS}) term by term in the above series to obtain the Fredholm determinant expansion of the superconformal index. One might worry about the exchange in the order of performing the Hubbard-Stratonovich transform and extracting the constant coefficient of the power series within the brackets $\{\dots\}$. However, since we're working with formal power series, it is easy to see that the two operations indeed commute, allowing us to write
\begin{align}
    Z_N\left(g_k\right)&=Z_\infty\left(g_k\right)\sum_{m=0}^\infty\frac{(-1)^m}{\left(m!\right)^2}[1]\Bigg\{\prod_{i=1}^m\left(\frac{u_i}{v_i}\right)^{N+1}\cdot\det\left(\frac{1}{1-u_i/v_j}\right)_{i,j=1,\dots,m}^2\cdot\nonumber\\
    &\cdot\exp\left[\sum_{k=1}^\infty\frac{1}{k}\frac{g_k}{\left(1-g_k\right)}\sum_{i=1}^m\left(u_i^k-v_i^k\right)\sum_{j=1}^m\left(v_j^{-k}-u_j^{-k}\right)\right]\Bigg\},\label{eqn:finalAppendixExpression}
\end{align}
where
\begin{equation}
    Z_\infty\left(g_k\right)\equiv\prod_{l=1}^\infty\frac{1}{1-g_l}.
\end{equation}
\section{Combinatorial Proof}\label{app:B}
The goal of this appendix is to provide a combinatorial proof of the equality between the expressions in equations (\ref{eqn:knownZkhat}) and (\ref{eqn:foundZkhat}),
\begin{equation}
    \sum_{m=1}^k\frac{1}{m!}\sum_{\substack{L_1,\dots,L_m\geq1\\L_1+\dots+L_m=k}}F_{L_1,\dots,L_m}^{(m)}(q)=\prod_{n=1}^k\frac{1}{1-q^n},
\end{equation}
where the terms $F_{L_1,\dots,L_m}^{(m)}(q)$ satisfy the following recurrence relation for $m\geq2$,
\begin{equation}
    F_{L_1,\dots,L_m}^{(m)}(q)=\frac{1}{1-q^{L_m}}F_{L_1,\dots,L_{m-1}}^{(m-1)}(q)-\sum_{j=1}^{m-1}F_{L_1,\dots,L_{j-1},L_j+L_m,L_{j+1},\dots,L_{m-1}}^{(m-1)}(q),\label{eqn:B2}
\end{equation}
with
\begin{equation}
    F_L^{(1)}(q)=\frac{1}{1-q^L}.
\end{equation}
We are grateful to Dongryul Kim for providing this proof and allowing us to include it in our work \cite{Daniel}.\\\\
The first step is to realize that the terms $F_{L_1,\dots, L_m}^{(m)}(q)$ can be written as
\begin{equation}
    F_{L_1,\dots,L_m}^{(m)}(q)=\sum_{\sigma\in S_m}\sgn(\sigma)\prod_{\substack{{\mathrm{cycles \;}} c\\{\mathrm{ of \;}}\sigma}}\frac{1}{1-q^{\sum_{j\in c}L_j}}. \label{eqn:B4}
\end{equation}
This follows inductively from the recurrence relation (\ref{eqn:B2}) as follows: For a permutation $\sigma\in S_m$ with $\sigma(m)=m$, letting $\tau\in S_{m-1}$, $\tau(j)=\sigma(j)$, the product over cycles of $\sigma$ splits into a factor of $1/\left(1-q^{L_m}\right)$ and the product over cycles of $\tau$,
\begin{align}
    \sum_{\substack{\sigma\in S_m\\ \sigma(m)=m}}\sgn(\sigma)\prod_{\substack{{\mathrm{cycles \;}} c\\{\mathrm{ of \;}}\sigma}}\frac{1}{1-q^{\sum_{j\in c}L_j}}&=\frac{1}{1-q^{L_m}}\sum_{\tau\in S_{m-1}}\sgn(\tau)\prod_{\substack{{\mathrm{cycles \;}} c\\{\mathrm{ of \;}}\tau}}\frac{1}{1-q^{\sum_{j\in c}L_j}}\nonumber\\
    &=\frac{1}{1-q^{L_m}}F_{L_1,\dots,L_{m-1}}^{(m-1)}(q).\label{eqn:B5}
\end{align}
The remaining terms of (\ref{eqn:B4}), with $\sigma(m)\neq m$, can be represented as products over the cycles of associated permutations $\tau\in S_{m-1}$, with $\tau(j)=\sigma(j)$ for $j\neq\sigma^{-1}(m)$ and $\tau\left(\sigma^{-1}(m)\right)=\sigma(m)$, and with $L_j$ replaced by $L_j'=L_j+\delta_{j,\sigma(m)}L_m$. Since $\sgn(\tau)=-\sgn(\sigma)$,
\begin{align}
    \sum_{\substack{\sigma\in S_m\\ \sigma(m)\neq m}}\sgn(\sigma)\prod_{\substack{{\mathrm{cycles \;}} c\\{\mathrm{ of \;}}\sigma}}\frac{1}{1-q^{\sum_{j\in c}L_j}}&=-\sum_{\tau\in S_{m-1}}\sgn(\tau)\prod_{\substack{{\mathrm{cycles \;}} c\\{\mathrm{ of \;}}\tau}}\frac{1}{1-q^{\sum_{j\in c}L_j'}}\nonumber\\
    &=-\sum_{j=1}^{m-1}F_{L_1',\dots,L_{m-1}'}^{(m-1)}(q).\label{eqn:B6}
\end{align}
Combining (\ref{eqn:B5}) and (\ref{eqn:B6}), we recover (\ref{eqn:B2}), proving the induction step.\\\\
We can then write
\begin{align}
        F_{L_1,\dots,L_m}^{(m)}(q)&=\sum_{\sigma\in S_m}\sgn(\sigma)\prod_{\substack{{\mathrm{cycles \;}} c\\{\mathrm{ of \;}}\sigma}}\frac{1}{1-q^{\sum_{j\in c}L_j}}\nonumber\\
        &=\sum_{\sigma\in S_m}\sgn(\sigma)\prod_{\substack{{\mathrm{cycles \;}} c\\{\mathrm{of \;}}\sigma}}\left(\sum_{n=0}^\infty q^{\sum_{j\in c}nL_j}\right)\nonumber\\
        &=\sum_{\sigma\in S_m}\sgn(\sigma)\sum_{\substack{n_1,\dots,n_m\geq0\\n_j=n_{\sigma(j)}}}q^{L_1n_1+\dots+L_mn_m}\nonumber\\
        &=\sum_{n_1,\dots,n_m\geq0}q^{L_1n_1+\dots+L_mn_m}\sum_{\substack{\sigma\in S_m\\n_j=n_{\sigma(j)}}}\sgn(\sigma).
\end{align}
For fixed $n_1,\dots,n_m\geq0$, we can see that unless all $n_j$ are distinct,
\begin{equation}
    \sum_{\substack{\sigma\in S_m\\n_j=n_{\sigma(j)}}}\sgn(\sigma)=0.
\end{equation}
This follows by noting that the sum over permutations with nontrivial cycles will contain an equal number of positive-sign permutations and negative-sign permutations. Thus,
\begin{equation}
    F_{L_1,\dots,L_m}^{(m)}(q)=\sum_{\substack{0\leq n_1,\dots,n_m\\ n_i\mathrm{\;all\;distinct}}}q^{L_1n_1+\dots+L_mn_m}.
\end{equation}
It remains to show that
\begin{equation}
    \sum_{m=1}^k\frac{1}{m!}\sum_{\substack{L_1,\dots,L_m\geq1\\L_1+\dots+L_m=k}}\sum_{\substack{0\leq n_1,\dots,n_m\\ n_i\mathrm{\;all\;distinct}}}q^{L_1n_1+\dots+L_mn_m}=\prod_{n=1}^k\frac{1}{1-q^n}.
\end{equation}
Due to the symmetry in $L_j$, we are allowed to order the $n_j$ while picking up a factor of $m!$,
\begin{equation}
    \sum_{m=1}^k\frac{1}{m!}\sum_{\substack{L_1,\dots,L_m\geq1\\L_1+\dots+L_m=k}}F_{L_1,\dots,L_m}^{(m)}(q)=\sum_{m=1}^k\sum_{\substack{L_1,\dots,L_m\geq1\\L_1+\dots+L_m=k}}\sum_{0\leq n_1<\dots<n_m}q^{L_1n_1+\dots+L_mn_m}.
\end{equation}
For fixed $m$, fixed $0\leq n_1<\dots<n_m$ and fixed $L_1,\dots,L_m\geq1$ with $L_1+\dots+L_m=k$, we let \begin{align}
    t_1&=\dots=t_{L_1}=n_1\nonumber\\
    t_{L_1+1}&=\dots=t_{L_1+L_2}=n_2\nonumber\\
    &\vdots\nonumber\\
    t_{k-L_m+1}&=\dots=t_{k}=n_m,
\end{align}
and note that the sum over $m$, $L_j$, and $n_j$ can then be written as a sum over $t_j$ with $0\leq t_1\leq\dots\leq t_k$:
\begin{equation}
    \sum_{m=1}^k\frac{1}{m!}\sum_{\substack{L_1,\dots,L_m\geq1\\L_1+\dots+L_m=k}}F_{L_1,\dots,L_m}^{(m)}(q)=\sum_{0\leq t_1\leq\dots\leq t_k}q^{t_1+\dots+t_k}.
\end{equation}
Letting $s_1=t_1$, $s_j=t_j-t_{j-1}$ for $j\geq2$, we see that
\begin{equation}
    \sum_{0\leq t_1\leq\dots\leq t_k}q^{t_1+\dots+t_k}=\sum_{s_1,\dots,s_k\geq0}q^{ks_1+(k-1)s_2+\dots+s_k}=\prod_{n=1}^k\frac{1}{1-q^n},
\end{equation}
concluding the proof.
\acknowledgments

Research supported by the Shoucheng Zhang Graduate Fellowship. I thank Raghu Mahajan, Dongryul Kim, Chitraang Murdia, Luca Iliesiu, and Gauri Batra for valuable discussions.


\end{document}